\documentclass[a4paper,11pt]{article}
\usepackage{amsfonts}

\usepackage{graphicx}
\graphicspath{{Figures/}}
\usepackage{amsmath}
\usepackage{amssymb}
\usepackage{latexsym}
\usepackage[colorlinks]{hyperref}
\usepackage{color}
\usepackage{float}
\usepackage{cite}
\usepackage{makeidx}
\usepackage{colortbl}
\usepackage{csquotes}
\usepackage[squaren]{SIunits}
\usepackage[toc]{appendix} 

\usepackage[textfont={footnotesize,sf},labelfont={color=blue,bf,sf},labelsep=endash]{caption}
\usepackage[position=top,labelfont={color=blue,bf,sf}]{subfig}
\usepackage{bm}

\newcommand{\bra}{\begin{array}}
\newcommand{\era}{\end{array}}
\newcommand{\beq}{\begin{equation}}
\newcommand{\eeq}{\end{equation}}

\begin{document}
\begin{titlepage}
\setcounter{page}{1}
\renewcommand{\thefootnote}{\fnsymbol{footnote}}
\begin{flushright}
\end{flushright}
\vspace{5mm}
\begin{center}
{\Large \bf {   
 Dynamics of Non-Gaussian Entanglement of Two Magnetically \\
 Coupled Modes   }
}\\

\vspace{5mm}
{\bf Radouan Hab-arrih}$^{a}$,   {\bf Ahmed Jellal}$^{a,b}$\footnote{\sf a.jellal@ucd.ac.ma} and {{\bf Abdeldjalil Merdaci$^{c}$}}

\vspace{5mm}
{$^{a}$\em Laboratory of Theoretical Physics,  
Faculty of Sciences, Choua\"ib Doukkali University},\\
{\em PO Box 20, 24000 El Jadida, Morocco}

{$^{b}$\em Canadian Quantum  Research Center,
	204-3002 32 Ave Vernon, \\ BC V1T 2L7,  Canada}

{$^{c}$\em Facult\'e des Sciences, Universit\'e 20 Ao\^ut 1955 Skikda,\\ BP 26, Route El-Hadaiek 21000, Algeria}

\vspace{30mm}

\begin{abstract}

 This paper surveys the quantum entanglement of two coupled harmonic oscillators via  angular momentum generating a magnetic coupling $\omega_{c}$.  The corresponding Hamiltonian is diagonalized by using three canonical transformations and then the stationary wave function is obtained. Based on the Schmidt decomposition, we explicitly determine the Schmidt modes  $\lambda_{k}$ with $k\in\left\lbrace 0,1,\cdots,n+m\right\rbrace$, $n$ and $m$ being two quantum numbers associated to the two oscillators. By studying 
 the effect of the anisotropy $ R=\omega_{1}^{2}/\omega_{2}^{2} $, 
 $\omega_{c}$, asymmetry $ |n-m| $ and  dynamics on the entanglement, we summarize our  results   as follows. $ (i)- $ The entanglement becomes very large with the increase of  $ (n,m) $. $ (ii)- $ The sensistivity to $\omega_c$  depends on  $ (n,m) $ and  $R$. $ (iii)- $ The periodic revival of entanglement strongly depends on the physical parameters and quantum numbers.
\vspace{30mm}

\end{abstract}
\end{center}
\end{titlepage}

\section{Introduction}

The entanglement of quantum particles is a physical phenomenon and remains among the most amazing features of the quantum world \cite{1a}. It accounts for the inseparability of the wavefunction describing a system of $ n $ particles to the tensor product of the state of each particle, i.e. $ \psi(x_1,\cdots,x_n)\neq \psi(x_1)\otimes \cdots\otimes \psi(x_n)$. Moreover,
the entanglement is a fundamental resource of the quantum information processing and by which the quantum technologies can go beyond the classical protocols \cite{1,2,3}. In entanglement theory, the most important questions can be summarized as follows. $ (i)- $ How to generate a large entanglement between particles in  a given physical set up. $ (ii)- $ How to mathematically compute  the  corresponding entanglement content \cite{4,13}. As for $ (i)$, the generation of entangled particles can be done for example  with  spontaneous parametric 
down-conversion, nuclear or atomic sources \cite{4a,4b,4c}. Regarding $ (ii)$, it depends on the type of the quantum states  under consideration, that is continuous or discrete, bipartite or multipartite, Gaussian or non-Gaussian, pure or mixed \cite{11,11a}. In the case of pure continuous bipartite states, 
the quantification of entanglement can be faithfully done via the von Neumann entropy $ S_{v}=-\sum_{k}\lambda_{k}\ln(\lambda_{k}) $ with $\lambda_{k}$ being the Schmidt modes of the state at hand. Thereby, computing the modes $ \lambda_{k} $ is not an easy task  because it involves complex integrals \cite{4,13}.  

The entanglement of systems made of coupled harmonic oscillators becomes a very active domain of research.  
In this respect, the pairwise entanglement of the ground state is largely studied,  for instance  we refer to \cite{5,8,9,10}. However, the entanglement of excited states of coupled harmonic oscillators is less studied, which is due to the mathematical complexity of computing the Schmidt modes $\lambda_{k}$ \cite{11,12}. If the Schmidt modes are computed, then the quantification of entanglement can easily  be obtained through Schmidt parameter $ K=\left(\sum_{k} \lambda_{k}^{2}\right)^{-1} $ or von Neumann entropy. Recently, in a seminal works,  Makarov has exactly obtained the analytical expressions of the Schmidt modes of  two harmonic oscillators connected via position-position coupling type, .i.e. $J\hat{x}_1\hat{x}_2$ \cite{13}, as well as with a coupling velocity-position interaction type, i.e. $ \alpha x_{1}p_{2}$ \cite{4}. As a result, Makarov showed that in both cases the entanglement becomes   very important for large  quantum numbers and under some specific choices of physical parameters.

Motivated by the Makarov works\cite{4,13},  we address the problem of two harmonic oscillators connected via an angular momentum coupling type. As a matter of clarity, this kind of coupling term is extensively used in several studies, for instance trapping ions \cite{14,15,17} and quantum invariant theory \cite{16,18,19}. We digonalize the Hamiltonian by using three canonical transformations and give the Schmidt decomposition of the obtained stationary and non-stationary wavefunctions. In addition, we harness  the derived modes  to compute the entanglement content. As a result, we numerically investigate  the effect of the anisotropy, asymmetry, magnetic coupling and  dynamics on the entanglement. Our obtained results are important not only from a theoretical point of view but also they lead to study a  generation of entangled photons via magnetic coupled waveguide beam splitters\cite{26,27,28}.

The layout of the paper is given as follows. In {\bf{\color{red}Sec} {\ref{Sec2}}}, we define our model and diagonalize it by involving three canonical transformations. The normal modes together with the stationary wave function of Schr\"{o}dinger equation will be obtained. The Schmidt decomposition will be analyzed in {\bf{\color{red}Sec} {\ref{Sec3}}} where the Schmidt modes will be explicitly computed. In {\bf{\color{red}Sec} {\ref{Sec4}}},  we quantify the  entanglement via two pure state quantifiers that are von Neumann entropy $ S_{v} $ and Schmidt parameter $ K $. We analyze the effect of the asymmetry $ |n-m| $, anisotropy $ R $ and the mixing angle $\theta$ on the entanglement content. In {\bf{\color{red}Sec} {\ref{Sec5}}}, we discuss  the dynamics of entanglement by using the dynamical Schr\"{o}dinger equation. Finally, we conclude our work.

\section{Hamiltonian of two modes \label{Sec2}}
We consider two  coupled harmonic oscillators connected via a transversal magnetic field $ {\bf{B}}=B_{z}u_{z} $ generating a rotation described by
the angular momentum $L_z$. Our system can be described by the Hamiltonian 
\begin{eqnarray}
\hat{H}=\frac{\hat{p}_{1}^{2}}{2}+\frac{\hat{p}_{2}^{2}}{2}+\frac{1}{2}\omega_{1}^{2}\hat{x}_{1}^{2}+\frac{1}{2}\omega_{2}^{2}\hat{x}_{2}^{2}+\omega_{c}\hat{L}_{z}\label{ham1}
\end{eqnarray}
where $\omega_{c}$ stands for the coupling frequency and $ L_{z}= \hat{x}_{1}\hat{p}_{2}-\hat{x}_{2}\hat{p}_{1}$. To diagonalize \eqref{ham1}, we proceed by introducing the transformation 
\begin{eqnarray}
&& \hat{p}_{1}=\dfrac{\hat{\mathfrak{p}}_{1}+\hat{\mathfrak{p}}_{2}}{\sqrt{2}\epsilon},\qquad x_{1}=\frac{\epsilon}{\sqrt{2}}\left(\hat{q}_{1}+\hat{q}_{2}\right)
\\
&&p_{2}=\dfrac{\hat{q}_{2}-\hat{q}_{1}}{\sqrt{2}\gamma},\qquad x_{2}=\frac{\gamma}{\sqrt{2}}\left(\hat{\mathfrak{p}}_{1}-\hat{\mathfrak{p}}_{2}\right) 
\end{eqnarray}
 with the commutation relations $ \left[\hat{q}_{j},\hat{\frak{p}}_{k} \right]=i\delta_{jk} $ and $\left[\hat{q}_{j},q_{k} \right]=\left[\hat{\frak{p}}_{j},\hat{\frak{p}}_{k} \right]=0$. By choosing the involved parameters as $\epsilon=1$ and $\gamma=\omega_{2}^{-1}$, then after  substituting into (\ref{ham1}) we get the Hamiltonian
\begin{eqnarray}
	\hat{H}_{1}=\dfrac{\hat{\mathfrak{p}}_{1}^{2}}{2m_{+}}+\dfrac{\hat{\mathfrak{p}}_{2}^{2}}{2m_{-}}+\frac{1}{2}\Omega_{+}^{2}\hat{q}_{1}^{2}+\frac{1}{2}\Omega_{-}^{2}\hat{q}_{2}^{2}+J\hat{q}_{1}\hat{q}_{2}
\end{eqnarray}
where  we have set $\Omega_{\pm}^{2}=\tfrac{\omega_{1}^{2}+\omega_{2}^{2}}{2}\mp \omega_{c}\omega_{2}$, $J=\tfrac{\omega_{1}^{2}-\omega_{2}^{2}}{2}$ and $ m_{\pm}=\tfrac{1}{1\mp\tfrac{\omega_{c}}{\omega_{2}}}$. Note that $  H_{1} $ is decoupled when  both modes are in resonance and remain coupled  out of it. 
To simplify the Hamiltonian diagonalization,  we perform a second transformation by reducing the kinetic matrix to unity 
\begin{eqnarray}
(\hat{y}_{1}, \hat{\Pi}_{2})=\sqrt[4]{ \tfrac{m_{+}}{m_{-}}}(\hat{q}_{1},\hat{\mathfrak{p}}_{2}), \qquad (y_{2}, \Pi_{1})=\sqrt[4]{ \tfrac{m_{-}}{m_{+}}}(\hat{q}_{2},\hat{\mathfrak{p}}_{1})
\end{eqnarray}
and then obtain the Hamiltonian
\begin{eqnarray}
\hat{H}_{2}=\dfrac{\hat{\Pi}_{1}^{2}}{2m}+\dfrac{\hat{\Pi}_{2}^{2}}{2m}+\frac{1}{2}m\varpi_{+}^{2}\hat{y}_{1}^{2}+\frac{1}{2}m\varpi_{-}^{2}\hat{y}_{2}^{2}+J\hat{y}_{1}\hat{y}_{2} \label{ham6}
\end{eqnarray}
where $m=\sqrt{m_{+}m_{-}}$ and $\varpi_{\pm}=\Omega_{\pm}/\sqrt{m_{\pm}}$. Because of the last term in (\ref{ham6}), $ H_{2} $ is not decoupled and then 
a third transformation is required. It can be realized  by rotating our system 
\begin{eqnarray}
&&
\hat{Q}_{1}=\cos\left(\frac{\theta}{2}\right)\hat{y}_{1}+\sin\left(\frac{\theta}{2}\right)\hat{y}_{2},\qquad \hat{Q}_{2}=-\sin\left(\frac{\theta}{2}\right)\hat{y}_{1}+\cos\left(\frac{\theta}{2}\right)\hat{y}_{2}\\
&&
\hat{P}_{1}=\cos\left(\frac{\theta}{2}\right)\hat{\Pi}_{1}+\sin\left(\frac{\theta}{2}\right)\hat{\Pi}_{2},\qquad \hat{P}_{2}=-\sin\left(\frac{\theta}{2}\right)\hat{\Pi}_{1}+\cos\left(\frac{\theta}{2}\right)\hat{\Pi}_{2} 
\end{eqnarray}
with $ \left[\hat{Q}_{j},\hat{P}_{k} \right]=i\delta_{jk} $ and $\left[\hat{Q}_{j},\hat{Q}_{k} \right]=\left[\hat{P}_{j},\hat{P}_{k} \right]=0$. We conclude that for a mixing angle fulfilling the condition
\begin{eqnarray} 
\tan\theta=\frac{(1-R)\sqrt{1-r^{2}}}{(1+3R)r}\label{mix}
\end{eqnarray}
with  $r=\omega_{c}/\omega_{2}$ and $ R=\omega_{1}^{2}/\omega_{2}^{2}\label{theta}$,  a diagonalized Hamiltonian can be obtained
\begin{eqnarray}
\hat{H}_{3}=\frac{\hat{P}_{1}^{2}}{2m}+\frac{\hat{P}_{2}^{2}}{2m}+\frac{1}{2}m\sigma_{1}^{2}\hat{Q}_{1}^{2}+\frac{1}{2}m\sigma_{2}^{2}\hat{Q}_{2}^2
\end{eqnarray}
where the involved frequencies are given by 
\begin{eqnarray}
\sigma_{1,2}^{2}
= \tfrac{1}{2}(\omega_{1}^{2}+\omega_{2}^{2}+2\omega_{c}^{2})\pm \tfrac{1}{2}\sqrt{r^{2}(3\omega_{2}^{2}+\omega_{1}^{2})^{2}+(\omega_{1}^{2}-\omega_{2}^{2})^{2}}.
\end{eqnarray}
Now, it is easy to solve the eigenvalue equation to derive the energies\cite{24}
\begin{eqnarray}
\label{ener}
E_{n,m}=\sigma_{1}(n+1/2)+\sigma_{2}(m+1/2)
\end{eqnarray}
and the corresponding wave functions read as 
\begin{eqnarray}
\psi _{n,m}\left( Q_{1},Q_{2}\right)  &=&\left( \frac{%
	\varpi }{\pi }\right) ^{\frac{1}{2}}\frac{1}{\sqrt{%
		2^{n+m}n!m!}}e^{-\tfrac{\varpi }{2 }\left(
	e^{-\eta }Q_{1}^{2}+e^{\eta }Q_{2}^{2}\right) } H_{n}\left( \sqrt{\varpi e^{-\eta }}%
Q_{1}\right) H_{m}\left( \sqrt{\varpi e^{\eta }}Q_{2}\right) \label{wave}
\end{eqnarray}
where  $\varpi=\sqrt{\sigma_{1}\sigma_{2}}$ and  $e^{-\eta}=\sigma_{1}/\varpi $.
In terms of the old coordinates, (\ref{wave}) can be mapped as
\begin{eqnarray}
\psi_{n,m}(x_{1}, p_{2})&=&\left( \frac{%
	\varpi }{\pi }\right) ^{\frac{1}{2}}\frac{1}{\sqrt{%
		2^{n+m}n!m!}}e^{-\tfrac{\varpi }{2 }\left(e^{-\eta}(S_{11}x_{1}+S_{14}p_{2})^{2}+e^{\eta}(S_{21}x_{1}+S_{24}p_{2})^{2}
\right) }  \notag  \\
&&\times H_{n}\left( \sqrt{\varpi e^{-\eta} }%
(S_{11}x_{1}+S_{14}p_{2})\right) H_{m}\left( \sqrt{\varpi e^\eta}(S_{21}x_{1}+S_{24}p_{2})\right)\label{9}
\end{eqnarray}
and $S_{ij}$ are  matrix elements of $S$, see the  {\color{red}\sf Appendix} \ref{appe}. To  write the wave functions in term of $ x_{1} $ and $ x_{2} $, we use the following Fourier transform
\begin{eqnarray}
\psi_{n,m}(x_{1},x_{2})=\int \tfrac{dp_{2}}{\sqrt{2\pi}} e^{ip_{2}x_{2}}\psi(x_{1},p_{2}). \label{10}
\end{eqnarray}
We point out here that the calculation of (\ref{10}) is not an obvious task. Fortunately, its analytical expression is not needed to achieve our object as we will see in the forthcoming analysis.

\section{Schmidt decomposition \label{Sec3}}
To study  the  entanglement of our system, let us  proceed by applying  the Schmidt decomposition techniques. Indeed, we decompose the stationary wave function as  
\begin{equation}\label{schm}
\psi _{n,m}\left( x_{1},p_{2}\right)
= \sum_{l,k=0}^{\infty }A_{n,m}^{k,l}\varphi _{k}\left(
x_{1}\right) \phi _{l}\left( p_{2}\right) 
\end{equation}
such that $\varphi _{k}\left( x_{1}\right) $ and $\phi _{l}\left( p_{2}\right) $
 are the vectors of the Schmidt basis 
\begin{align}
\phi _{l}\left( x_{1}\right) & =\tfrac{1}{\sqrt{2^l l!}} \sqrt[4]{\tfrac{{\omega_{1} }}%
{{\pi }}} \, e^{-\tfrac{\omega_{1}}{2} %
	x_{1}^{2}} H_{l}\left( \sqrt{\omega_{1} }x_{1}\right)\\
\varphi _{k}\left( p_{2}\right)  & =\tfrac{1}{\sqrt{2^k k!}} \tfrac{1}{\sqrt[4]{\pi \omega_{2}} }\,e^{-\tfrac{1}{2\omega_{2}} %
	p_{2}^{2}}H_{k}\left( \tfrac{p_{2}}{\sqrt{\omega_{2} }}\right)
\end{align}
which
correspond to the unbound oscillators and satisfy 
 $\left\langle \phi_{k} \vert \phi_{l} \right\rangle=\left\langle \varphi_{k} \vert \varphi_{l} \right\rangle=\delta_{kl}$. Consequently, the coefficients $A_{n,m}^{k,l}$ will be  obtained by using the
orthogonality properties%
\begin{equation}
A_{n,m}^{k,l}=\iint dx_{1}dp_{2} \psi _{n,m}\left(
x_{1},p_{2}\right) \varphi _{l}\left( x_{1}\right) \phi _{k}\left(
p_{2}\right)\label{decomposition}.
\end{equation}%
This can be  computed by employing  the Rodrigues formula 
\begin{equation}
H_{n}\left( \omega x\right) =\frac{d^{n}}{du^{n}}\left. e^{-u^{2}+2\omega
	xu}\right\vert _{u=0}
\end{equation}%
to get the result 
\begin{eqnarray}
A_{n,m}^{k,l}&=& \tfrac{1}{\pi}\sqrt{\frac{\varpi \sqrt{\omega_{1}}}{ 
			\sqrt{\omega_{2}}2^{n+m+k+l}n!m!k!l!}} \iint \tfrac{d^{k}}{ds^{k}} \tfrac{d^{l}}{dw^{l}}\tfrac{d^{m}}{dv^{m}}\tfrac{d^{n}}{du^{n}}
			\Big[  e^{-\tfrac{\varpi }{2 }\left(e^{-\eta}(S_{11}x_{1}+S_{14}p_{2})^{2}+e^{\eta}(S_{21}x_{1}+S_{24}p_{2})^{2}
	\right) }  \notag  \\
&& e^{-\tfrac{1}{2\omega_{2}} 
	p_{2}^{2}-\tfrac{\omega_{1}}{2} %
	x_{1}^{2}}\ e^{-u^{2}+2\sqrt{\varpi e^{-\eta}}%
	(S_{11}x_{1}+S_{14}p_{2})u}\ e^{-v^{2}+2\sqrt{\varpi e^{\eta}}%
	(S_{21}x_{1}+S_{24}p_{2})v}\notag\\
&& e^{-w^{2}+2(\sqrt{\omega_{1} } x_{1}w)} \ e^{-s^{2}+\frac{2}{\sqrt{\omega_{2} }} p_{2}s} \Big] \Big\vert_{u,v,w,s=0}  dx_{1}dp_ {2} \label{29}.
\end{eqnarray}
To proceed further, we consider an assumption based on  the fact that the magnetic coupling  $\omega_{c}$ is very small compared to the transition frequencies $\omega_{1,2}$, i.e. $\omega_{c}\ll\min(\omega_{1}, \omega_{2})$ \cite{4,13,26,27}. Consequently, the masses reduce to $ m_{\pm}\sim 1 $. Besides, we assume that $\omega_{1}$ and $\omega_{2}$ are very close such that their difference approaches to $\omega_c$, i.e. 
$ |\omega_{1}-\omega_{2}|\sim \omega_{c}$, and the normal frequencies behave like $\omega_{1}\sim \omega_{2}\sim\sigma_{1}\sim\sigma_{2}$ resulted in having  the phase $ \eta\sim0 $.  It is worthy to mention also that taking these approximations into account does not prevent saying that the  mixing angle $ \theta $ is free and can take any values in $ ]-\pi/2,\pi/2[ $. At this stage, one can show that the integral (\ref{29})
becomes 
\begin{eqnarray}
A_{n,m}^{k,l}&=&\tfrac{1}{\pi}\sqrt{\frac{\varpi}{ 
			2^{n+m+k+l}n!m!k!l!}} \iint \tfrac{d^{k}}{ds^{k}} \tfrac{d^{l}}{dw^{l}}\tfrac{d^{m}}{dv^{m}}\tfrac{d^{n}}{du^{n}}\Big[  e^{-\varpi
	(x_{1}^{2}+\frac{1}{\varpi^{2}}p_{2}^{2})  }  \notag  \\
&& e^{-u^{2}+2\sqrt{\varpi e^{-\eta}}%
	(S_{11}x_{1}+S_{14}p_{2})u}\ e^{-v^{2}+2\sqrt{\varpi e^{\eta}}%
	(S_{21}x_{1}+S_{24}p_{2})v}\notag\\
&&  e^{-w^{2}+2(\sqrt{\varpi } x_{1}w)}\ e^{-s^{2}+\frac{2}{\sqrt{\varpi }} p_{2}s} \Big] \Big|_{u,v,w,s=0}  dx_{1}dp_ {2}.
\end{eqnarray}
Performing the integration over $ x_{1} $  and $ p_{2} $, we end up with
\begin{eqnarray}
A_{n,m}^{k,l}&=& \frac{%
1 }{{\sqrt{%
			2^{n+m+k+l}n!m!k!l!}}} \tfrac{d^{k}}{ds^{k}} \tfrac{d^{l}}{dw^{l}}\tfrac{d^{m}}{dv^{m}}\tfrac{d^{n}}{du^{n}}\left[ e^{2u(S_{11}w+\varpi S_{14}s)+2v(S_{21}w+\varpi S_{24}s)}\right]\Big|_{u,v,w,s=0}.
\end{eqnarray}
Using the fact that $\tfrac{d^{n}}{du^{n}}u^{k}\vert_{u=0}=k!\delta_{k,n}$ and the following identity \cite{25}
\begin{eqnarray}
\tfrac{d^{n}}{du^{n}}\left[u^{\alpha}(a-u)^{\beta} \right] =n!a^{n}u^{\alpha-n}(a-u)^{\beta-n}P_{n}^{(\alpha-n,\beta-n)}\left(1-\tfrac{2u}{a}\right) 
\end{eqnarray}
we obtain
\begin{eqnarray}
A_{n,m}^{k,l}
		= (-1)^{m}%
			\sqrt{\tfrac{2^{n+m}} {
					2^{k+l}n!m!k!l!}}S_{11}^{n}S_{21}^{m} \tfrac{d^{k}}{ds^{k}}\tfrac{d^{l}}{dw^{l}}\left[\left (w+\tfrac{S_{21}}{S_{11}}s\right)^{n}\left(\tfrac{2}{\cos\theta}s-w-\tfrac{S_{21}}{S_{11}}s\right)^{m}\right]  \Big\vert _{w,s=0}.
				\end{eqnarray}
				{Let us introduce the change 
			 $ W=w+\tfrac{S_{21}}{S_{11}}s $ and then we get
			 \begin{eqnarray}
			 A_{n,m}^{k,l}
	 		 =(-1)^{m}
	 		 	\sqrt{\tfrac{l!k!} {
	 		 			2^{k-l}n!m!}}\ P_{l}^{(n-l,m-l)}\left( \sin\theta\right)(1-\sin\theta)^{\tfrac{n-l}{2}}(1+\sin\theta)^{\tfrac{m-l}{2}}\delta_{n+m-l,k} \label{conservation}
			 \end{eqnarray}
where $ P^{(\alpha,\beta)}_{n}(x) $ stands for Jacobi polynomials. Using the conservation formula of indexes 
$
n+m=k+l
$
one can rewrite $ A_{n,m}^{l,k} $ as 
\begin{eqnarray}
A_{n,m}^{l,n+m-l}=(-1)^{m}\sqrt{\frac{
	l!(n+m-l)! }{%
			2^{n+m-2l}n!m!}}\ P_{l}^{(n-l,m-l)}\left( \sin\theta\right)(1-\sin\theta)^{\tfrac{n-l}{2}}(1+\sin\theta)^{\tfrac{m-l}{2}}.
\end{eqnarray}
Now, it is  time to discuss the physical meaning of the coefficients  $  A_{n,m}^{k,l} $. Indeed, if the the system is initially in the state $|k,l\rangle$ it will be detected in the state $|p_{1},p_{2}\rangle$ with the probability \cite{4,13} 
\begin{eqnarray}
a_{p_{1},p_{2}}(t)=\sum\limits_{n=0}^{k+l}A^{n,k+l-n}_{k,l} A^{\ast n,k+l-n}_{ p_{1},p_{2}}e^{-i\Delta E_{n,k+l-n}}
\end{eqnarray}
such that the energy variation reads as 
\begin{eqnarray}
  \Delta E_{n,k+l-n}= E_{n,k+l-n}-\omega_{1}(k+\tfrac{1}{2})-\omega_{2}(l+\tfrac{1}{2}).
\end{eqnarray}
and $ E_{n,k+l-n} $ is given in \eqref{ener}. In the next, we will see how the above results will be employed to discuss the entanglement of our system.

\section{Quantum entanglement \label{Sec4}}
\subsection{Entanglement quantifiers}

Two quantifiers can be used to study the entanglement and to do one needs to involves the reduced matrices. Then according to our  Schmidt decomposition
\eqref{schm}, 
we straightforwardly derive the required densities as
\begin{eqnarray}
\rho_{1}(x_{1},x^{'}_{1},t)&=&\sum\limits_{l=0}^{m+n}|A_{n,m}^{l,n+m-l}|^{2}\varphi_{l}(x_{1})\varphi^{\ast}_{l}(x_{1}^{'})\\
\rho_{2}(p_{2},p_{2}^{'},t)&=&\sum\limits_{l=0}^{m+n}|A_{n,m}^{l,n+m-l}|^{2}\phi_{m+n-l}(p_{2})\phi^{\ast}_{m+n-l}(p_{2}^{'})
\end{eqnarray}
 where  $ \varphi_{l} $ and $ \phi_{m+n-l} $ are vectors of the Schmidt basis,
with $n+m $ is the Schmidt number. From the above considerations, one can derive the Schmidt modes as 
\begin{eqnarray}
\lambda_{k}&=&|A_{n,m}^{l,n+m-l}|^{2}.
\end{eqnarray}
Consequently, quantifying entanglement becomes a simple task. Indeed, this can be done using the von Neumann entropy $ S_{v} $ or the Schmidt parameter $ K $ defined as \cite{12,28}
\begin{eqnarray}
S_{v}=-\sum_{k=0}^{m+n}\lambda_{k}\ln(\lambda_{k}), \qquad K=\left( \sum_{k=0}^{m+n} \lambda_{k}^{2}\right)^{-1}.
\end{eqnarray}
These results will be used to show  the asymmetry, anisotropy and magnetic coupling effects on the entanglement.

\subsection{Asymmetry and anisotropy effect}

It should be noted here that the study of entanglement can be performed only by using the angle $ \theta $ given in (\ref{mix}). In {\bf \color{blue}\sf{Figure}} \ref{fig1}, we plot both quantifiers ($S_{v}$ and $K$) versus $ \sin\theta $ for various quantum numbers $(n,m)$. We  observe  that  the entanglement strongly depends on the quantum pairs $ (n,m) $ because  its hierarchy 
 is clear since it increases  by increasing   $ (n,m) $. Subsequently as expected,  the entanglement disappears for $\theta\rightarrow\pm\pi/2$, which corresponds to uncoupled of the two modes,  i.e. $\omega_{c}=0$, with  $\pm$ reflects the sign of frequency $(\omega_{1}-\omega_{2})$. Additionally for different quantum numbers $ n\neq m $,  the entanglement  reaches maximal values for $\theta=0$, which tells us that our system is isotropic in this case, i.e. 
 $\omega_{1}=\omega_{2}$.

\begin{figure}[H]
	\centering
	\includegraphics[width=8cm, height=6.3cm]{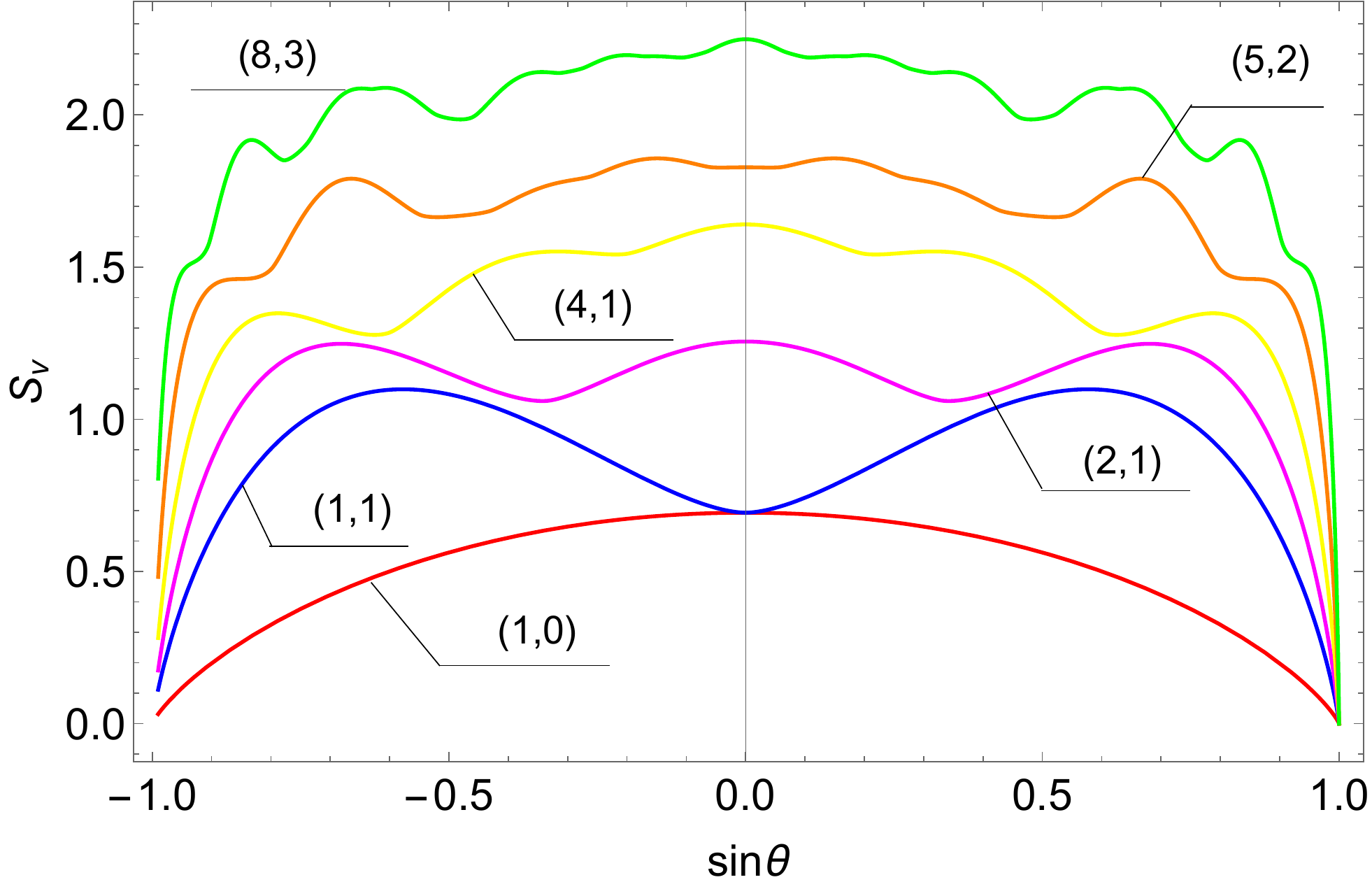}\ \ \ \ \ \
	\includegraphics[width=8cm, height=6.3cm]{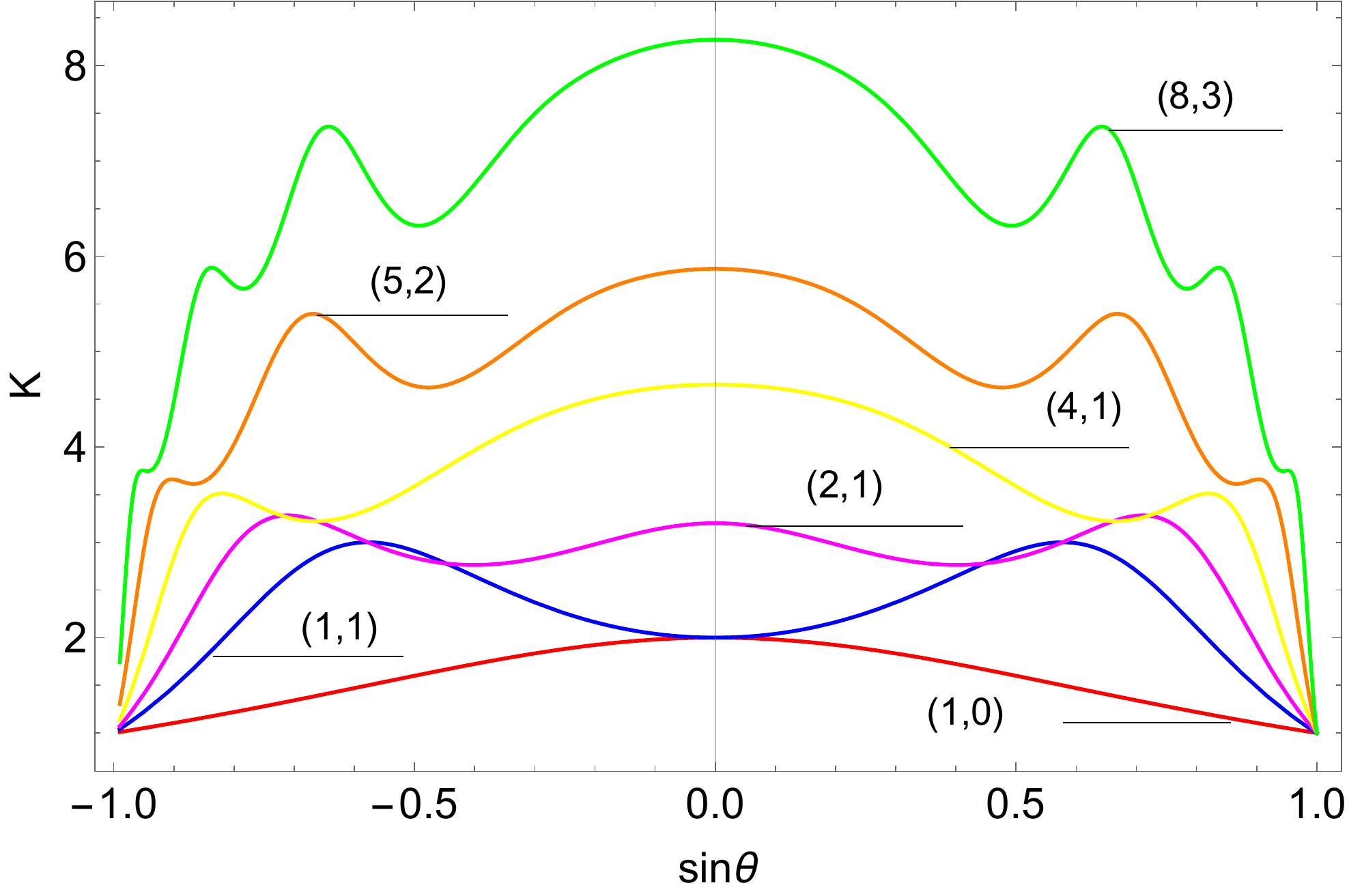}
	\captionof{figure}{\sf (color online)  The quantum entanglement via von Neumann entropy $S_{v}$ and Schmidt parameter $K$  versus the mixing angle $ \sin\theta$. The pairs stand for the quantum numbers $(n,m)$.\label{fig1}}
\end{figure}

 To show the entanglement when both oscillators have the same state $ n=m $, we plot in (left panel) of {\bf \color{blue}\sf{Figure}} \ref{fig2} the entanglement versus $\sin\theta$  for various pairs with $ (n,n) $. The optimal values of the entanglement is not obtained for the isotropic case, but it depends strongly on the quantum number $  n $ and approaches $\theta=0$ as $ n $ increases. As example, for $n=2$, the optimal value of entanglement is obtained for $ \sin\theta_{op}\sim 0.4  $, while  for $ n=3 $ we have $ \sin\theta_{op}\sim 0.3 $. In the (right panel) of {\bf \color{blue}\sf{Figure}} \ref{fig2}, we show the effect of the state asymmetry defined as $|n-m|$. Assuming for instance that $ n+m=10 $ and varying $ n $ and $ m $, we observe that the entanglement has not a monotically behavior. However, it becomes more important for the symmetry case 
 $ n=m $ in the full range of $ \theta $ except at vicinity of the isotropic regime. Now, by increasing the asymmetry $ |n-m| $, the entanglement in the vicinity of isotropic regime for non-symmetric state  becomes more important than that of symmetric ones. 
\begin{figure}[H]
	\centering
	\includegraphics[width=8cm, height=6.3cm]{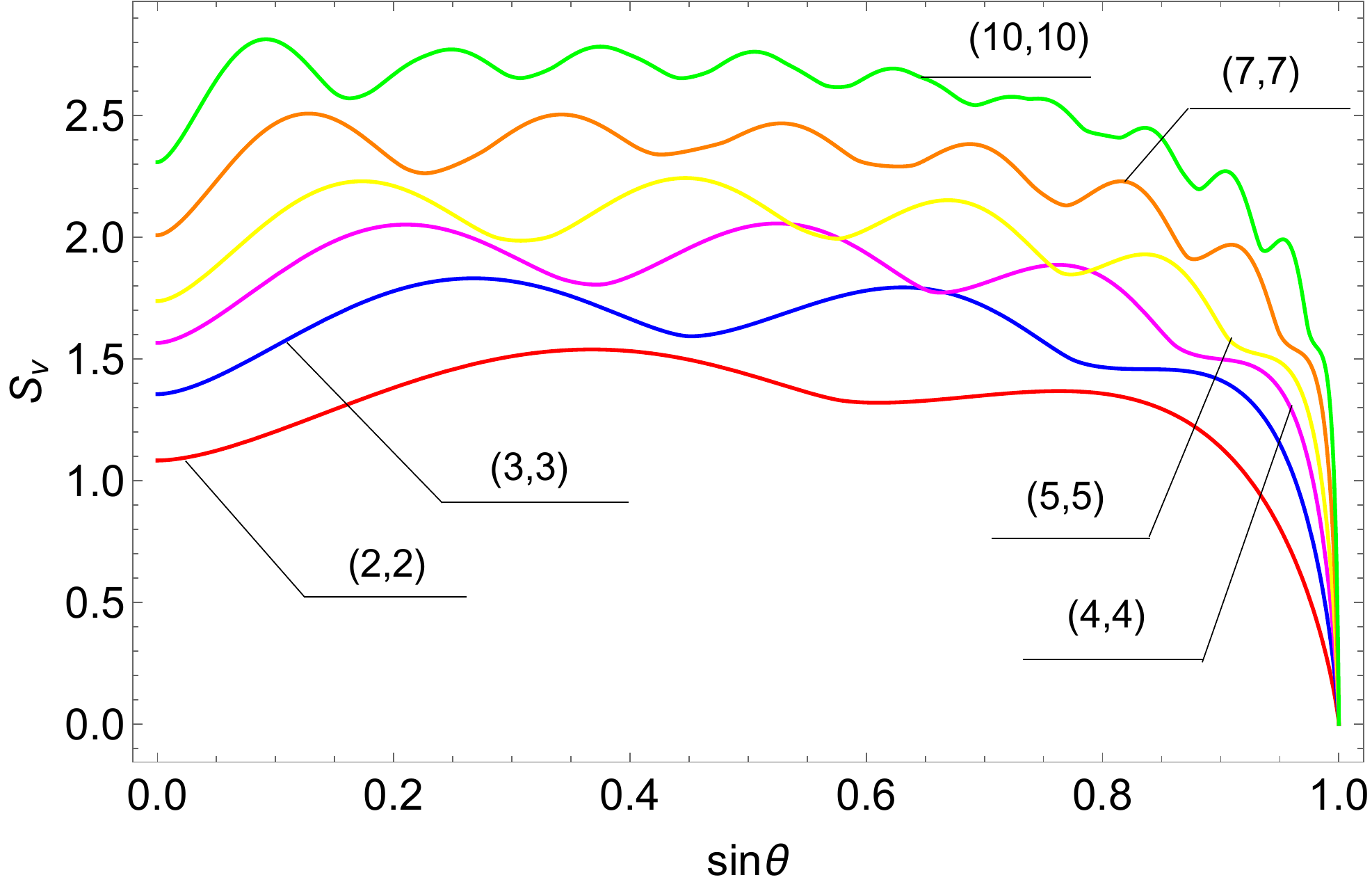}\ \ \ \ \ \ \
	\includegraphics[width=8cm, height=6.3cm]{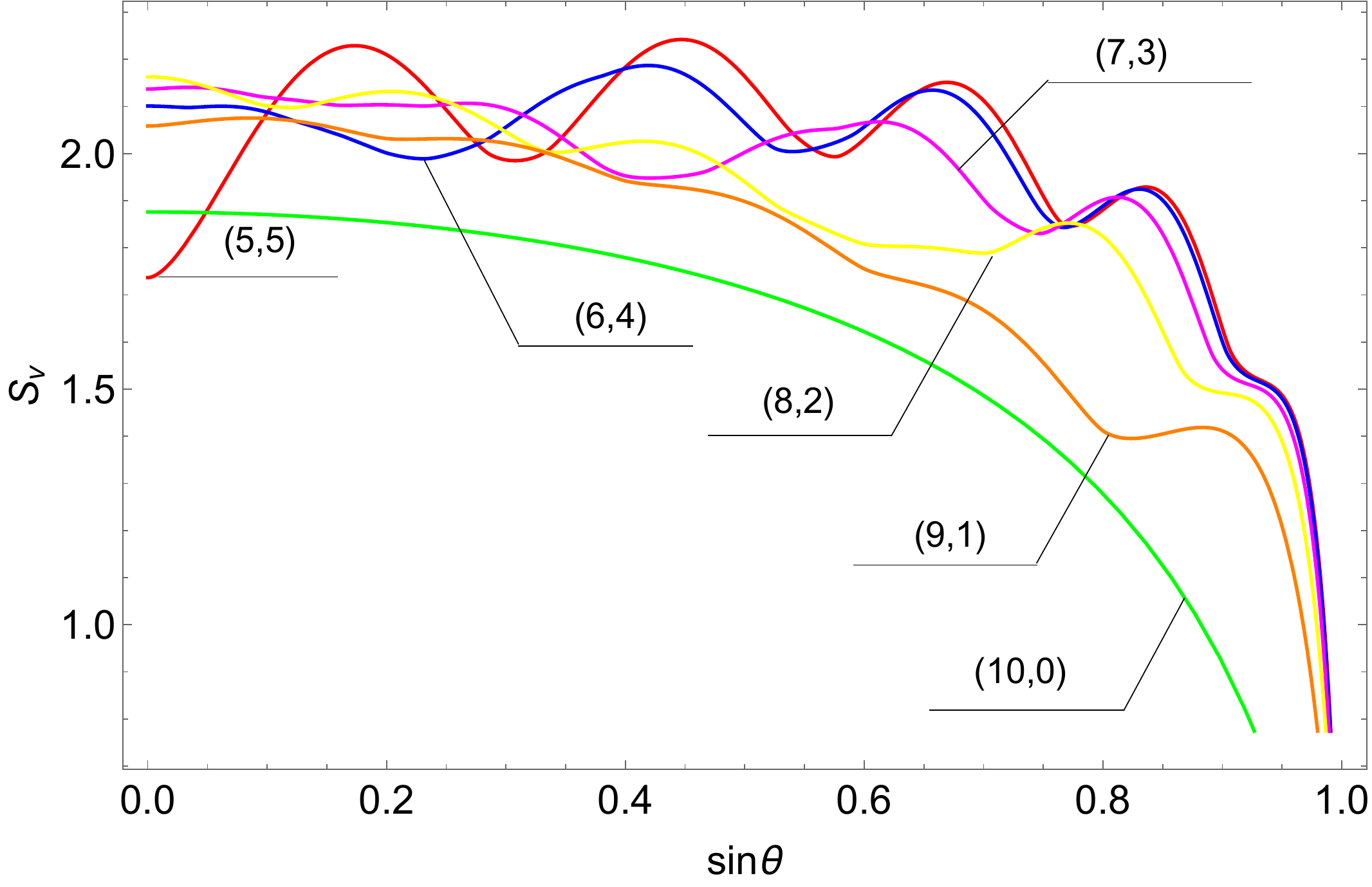}
	\captionof{figure}{\sf (color online)  The quantum entanglement  versus $\sin\theta$ with the same states in left panel and $n+m=10$ in right panel.\label{fig2}}
\end{figure}

\subsection{Magnetic coupling effect}
In {\bf \color{blue}\sf{Figure}} \ref{fig3}, we present the effect of the magnetic coupling $r=\tfrac{\omega_{c}}{\omega_{2}}$  on the entanglement. In panels  (a,b) we choose $ R=\tfrac{\omega_{1}^{2}}{\omega_{2}^{2}}=0.97 $, as observed after its revival, the entanglement is frozen after a fast oscillatory behavior. This indicates that the entanglement of  large quantum numbers is  more sensitive to the coupling $r$. For the maximal asymmetric states $ |n-m|=\max(m,n)$, the entanglement revives exponentially and will be  immediately frozen, which indicates that for these states the entanglement is indifferent to $r$. Additionally, to show the  magnetic coupling effect in the vicinity of resonance, i.e. $R\rightarrow 1$, we plot in the panels (c,d)  the case for $ R=0.999 $. Consequently, we notice that the sensitivity to $r$  decreases and after its revival, the entanglement will be rapidly frozen.
\begin{figure}[H]
	\centering
	\includegraphics[width=7cm, height=5.5cm]{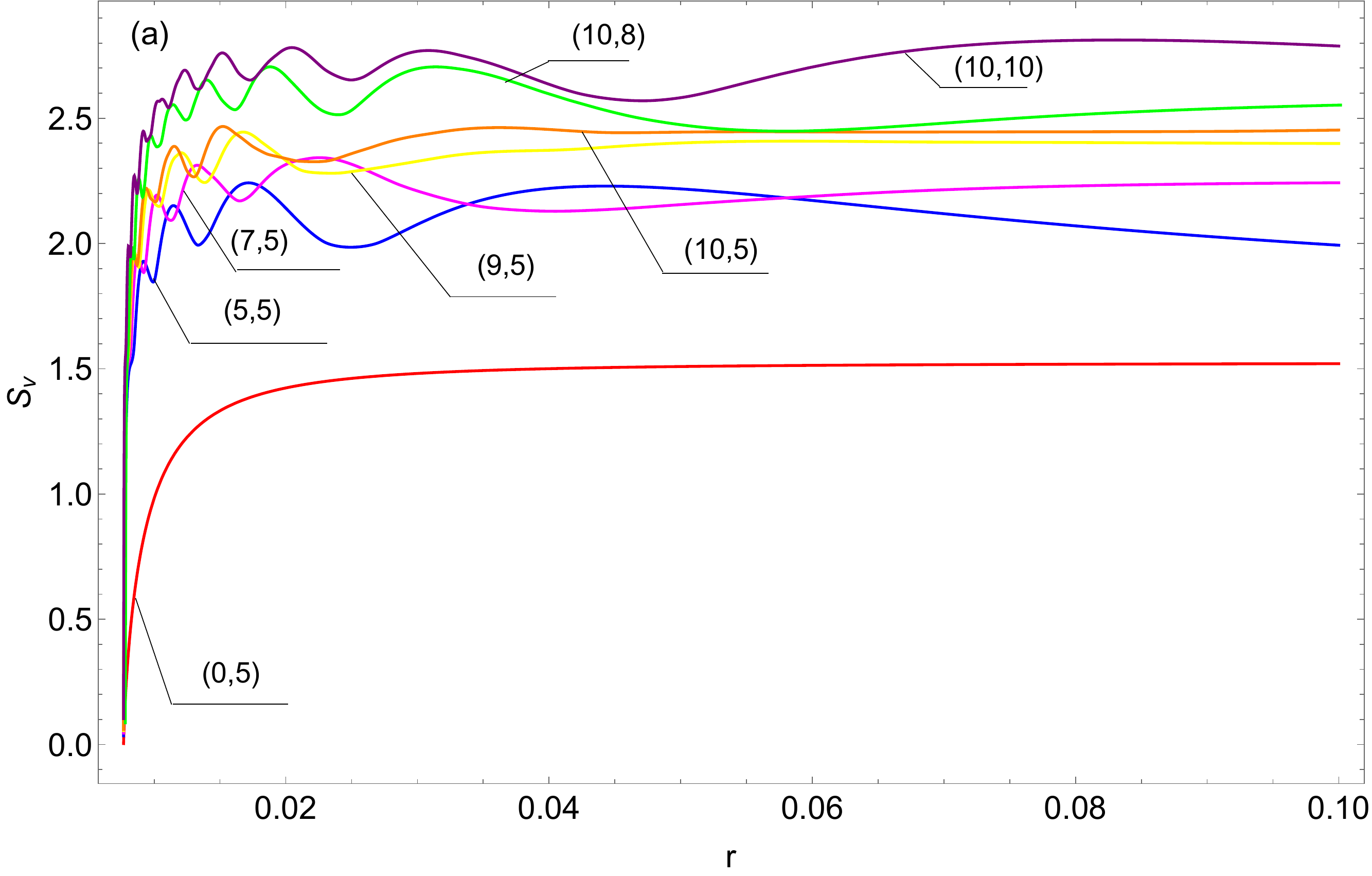}\ \ \ \ \ \
	\includegraphics[width=7cm, height=5.5cm]{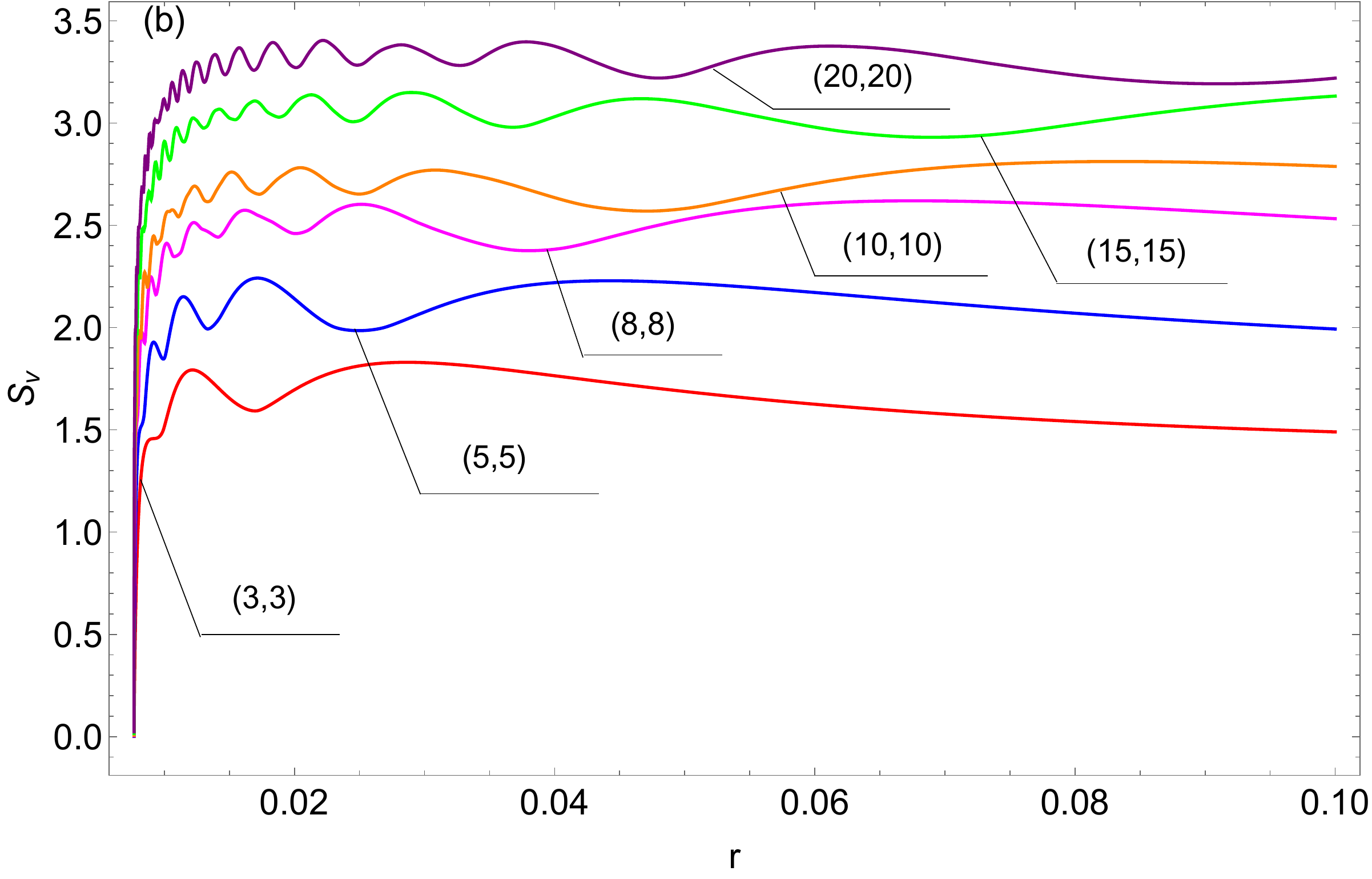}
	\includegraphics[width=7cm, height=5.5cm]{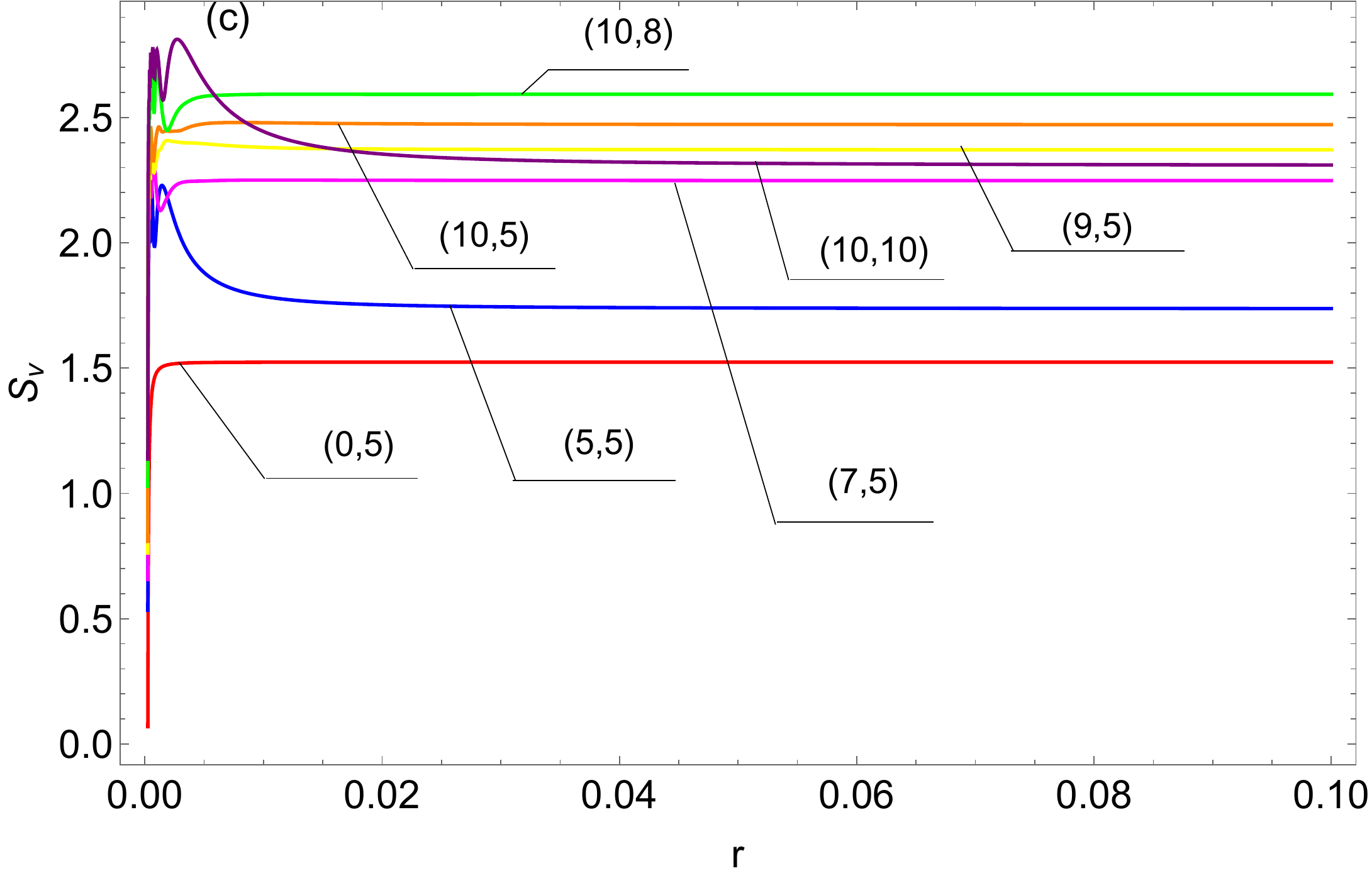}\ \ \ \ \ \
	\includegraphics[width=7cm, height=5.5cm]{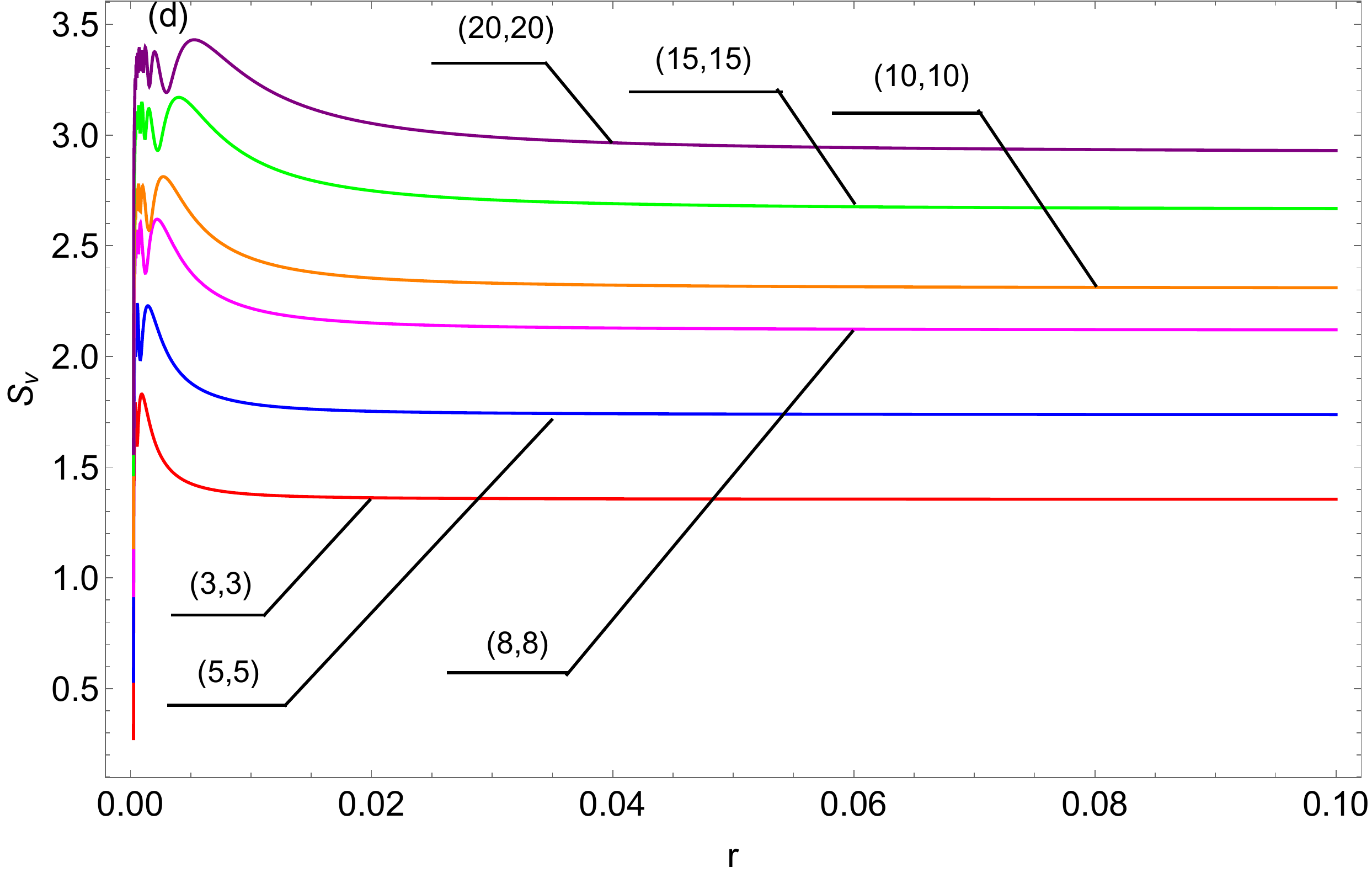}
	\captionof{figure}{\sf (color online)  The quantum entanglement  versus the magnetic coupling $r=\omega_{c}/\omega_{2}$ for different states $(n,m)$. Left panels for $R=0.97$ and  right panel for $R=0.999$. \label{fig3}}
\end{figure}

\section{Dynamics of entanglement \label{Sec5}}
Now, we consider the dynamical Schr\"{o}dinger equation associated with the Hamiltonian (\ref{ham1}). Initially, we assume both oscillators are separable $(\omega_{c}=0)$ and having the quantum states 
$|m_{1}\rangle$, $|m_{2}\rangle$, i.e. the wave functions of the uncoupled oscillators. In order to quantify the entanglement encoded in our state, we reapply the Schmidt decomposition of the non-stationary Schr\"{o}dinger equation. The expansion of the wave function is 
\begin{eqnarray}
\psi(x_1,x_2,t)&=&\sum\limits_{k,m} d_{k,p}(t)\phi_{k}(x_{1},t)\varphi_{p}(p_{2},t)
\end{eqnarray} 
where  $ \phi_{k}(x_1,t) $ and $\varphi_{p}(p_2,t)$ are the wave functions related to the unbound oscillators. 
The time-dependent expansion coefficients $ d_{k,p}(t)$  can be computed  by using  the integral form (\ref{decomposition}) to get
\begin{eqnarray}
d_{k,p}(t)&=& \sum\limits_{m=0}^{m_1+m_2}A^{m,m_{1}+m_{2}-m}_{m_{1},m_{2}}A^{\star m,m_{1}+m_{2}-m}_{k,p}e^{-i\Delta E_{m,m_1+m_2-m}t} 
\end{eqnarray}
with  energy difference
\begin{eqnarray}
\Delta E_{m,m_1+m_2-m}&=&E_{m,m_1+m_2-m}-\omega_{1}(m_{1}+\tfrac{1}{2})-\omega_{2}(m_{2}+\tfrac{1}{2}).
\end{eqnarray}
By virtue of (\ref{conservation}), we find new conservation formula $ k+p=m_{1}+m_{2} $. 
By making use of the reduced density matrices, we straightforward  end up with the time-dependent Schmidt modes 
\begin{eqnarray}
\lambda_{k}(t)&=& |d_{k,m_{1}+m_{2}-k}(t)|^{2}
\end{eqnarray}
Under the assumption of the weak coupling, i.e. $ r\ll 1$, and  $\omega_{1}\sim \omega_{2}$, one can show that 
the energy difference depends only on the quantum number $m$
\begin{equation}
 \Delta E_{m,m_1+m_2-m}t \sim \epsilon m t
\end{equation}
with $\epsilon=\sigma_{1}-\sigma_{2}$. In what follows, we will use the dimensionless time $\tilde{t}=\epsilon t$ to study the dynamics of entanglement.

\begin{figure}[H]
	\centering
	\includegraphics[width=6cm, height=5cm]{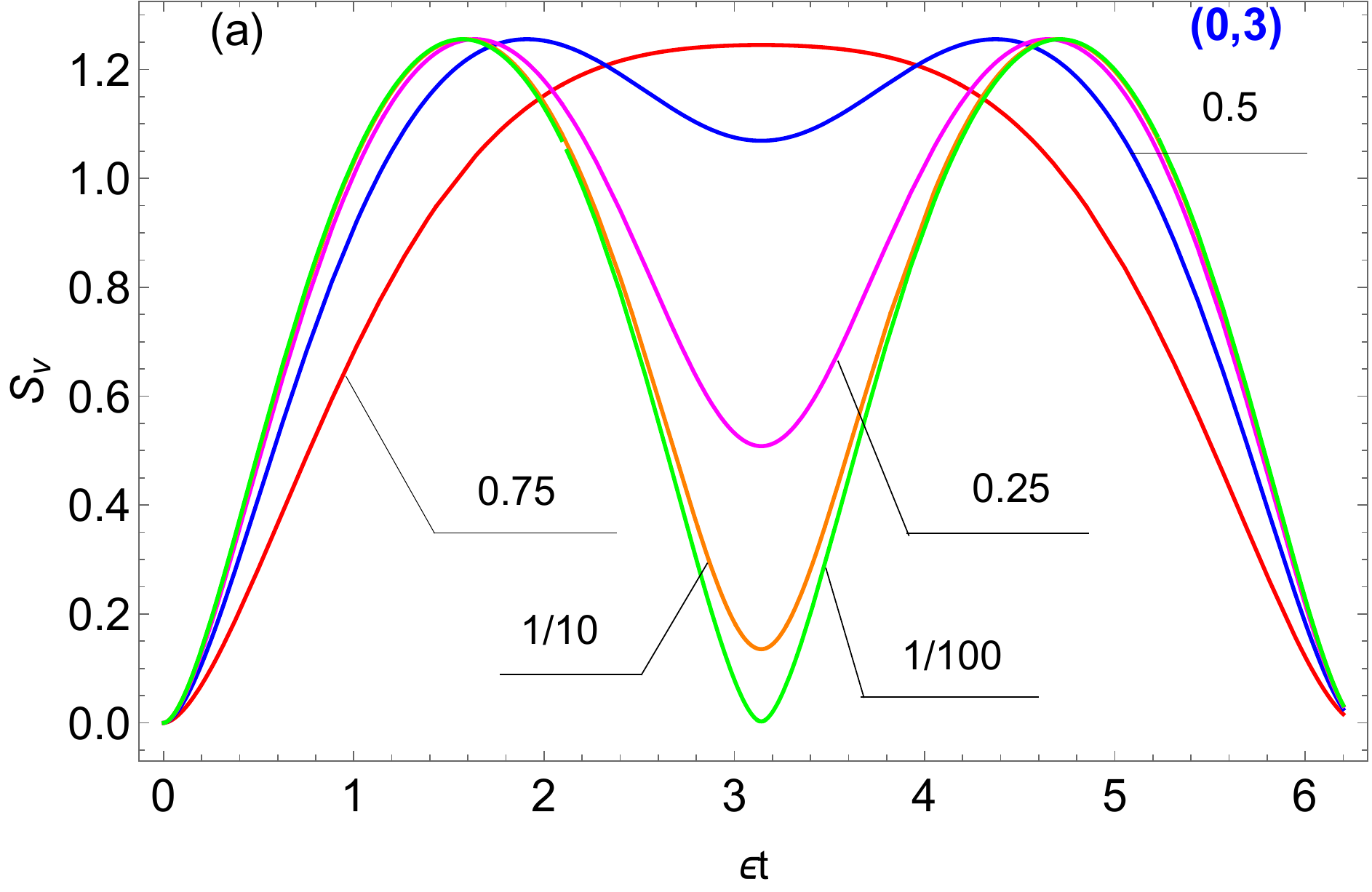}\ \ \ \ \ \
		\includegraphics[width=6cm, height=5cm]{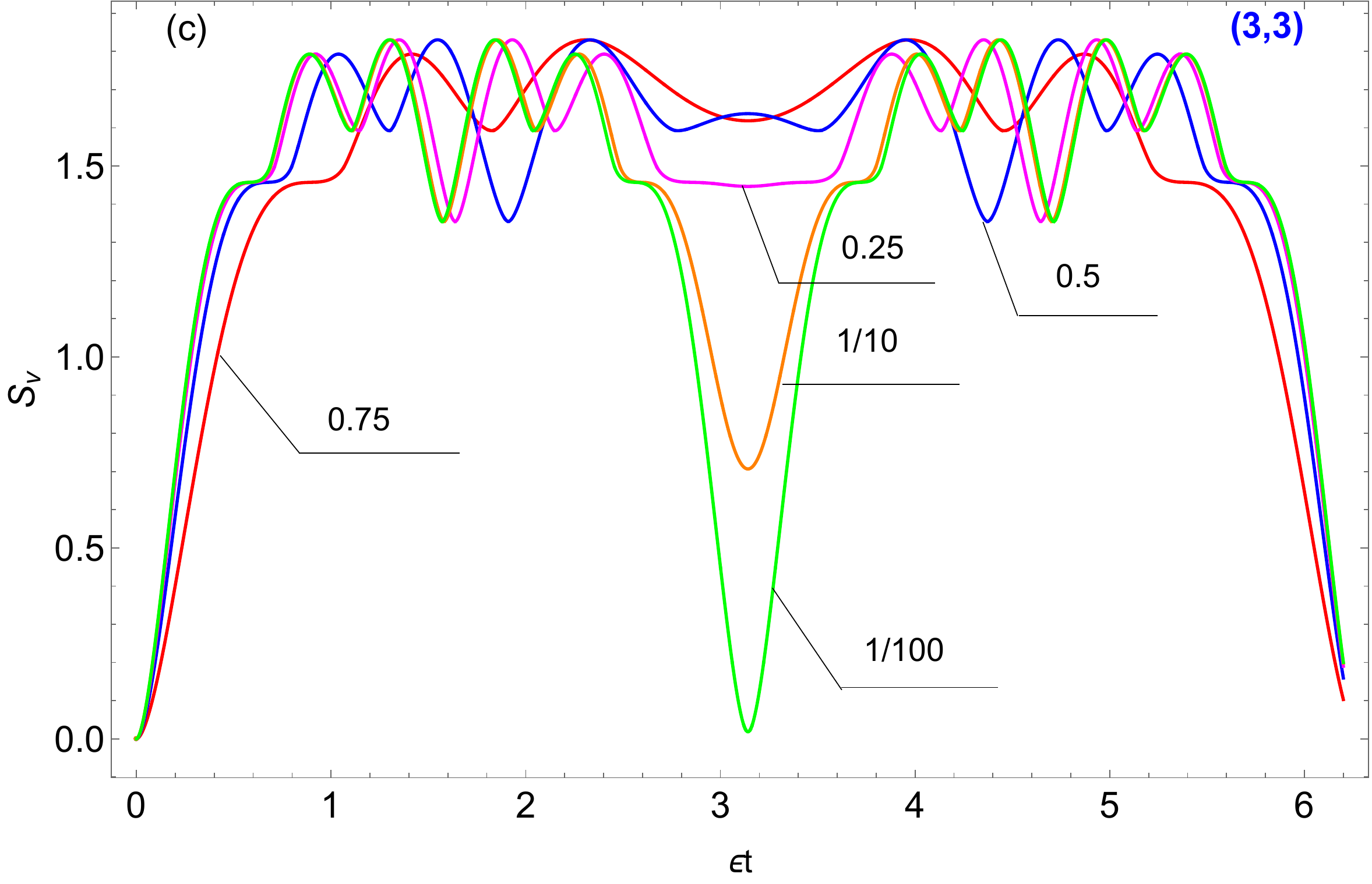}
		\includegraphics[width=6cm, height=5cm]{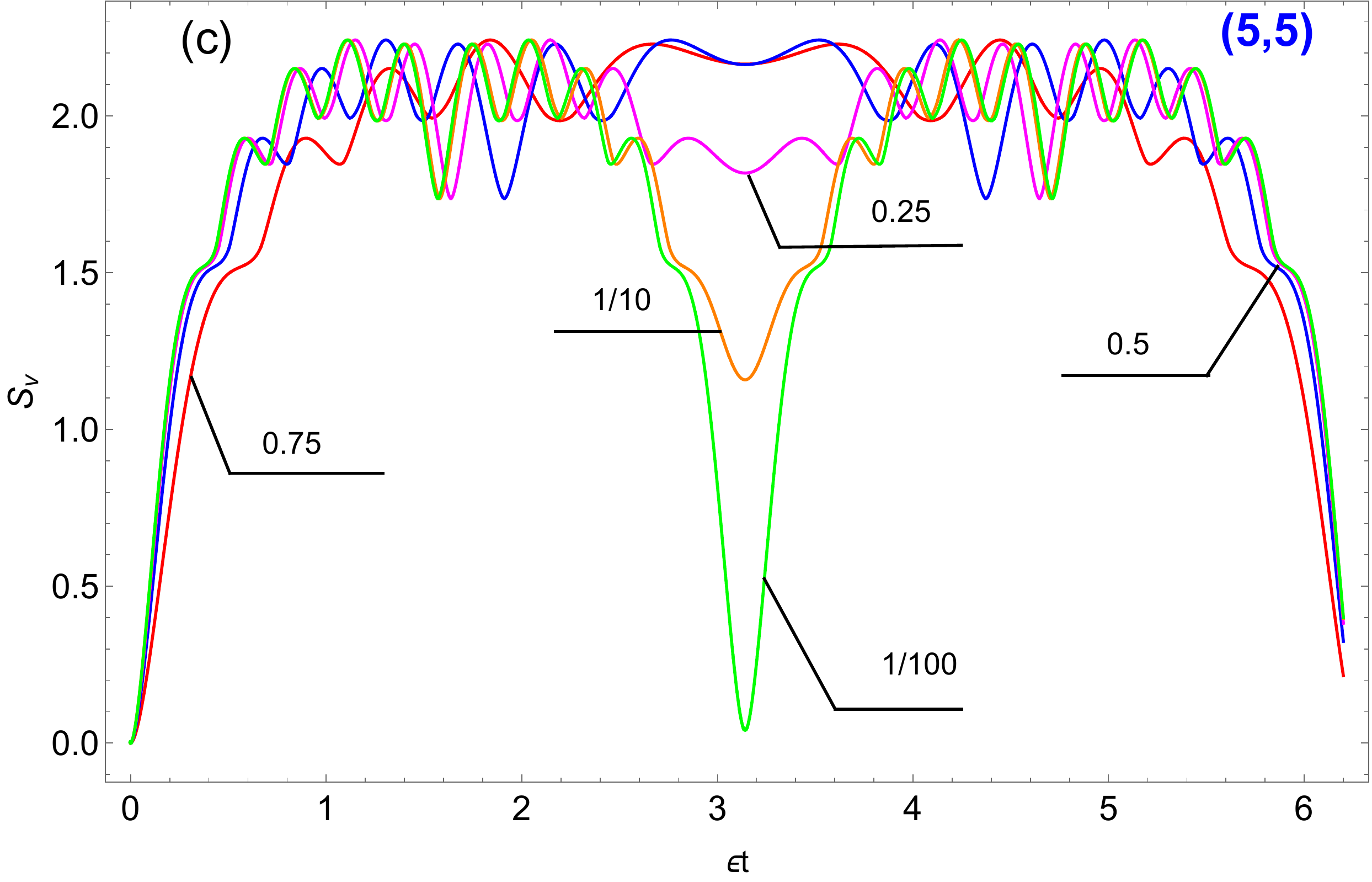}\ \ \ \ \ \
		\includegraphics[width=6cm, height=5cm]{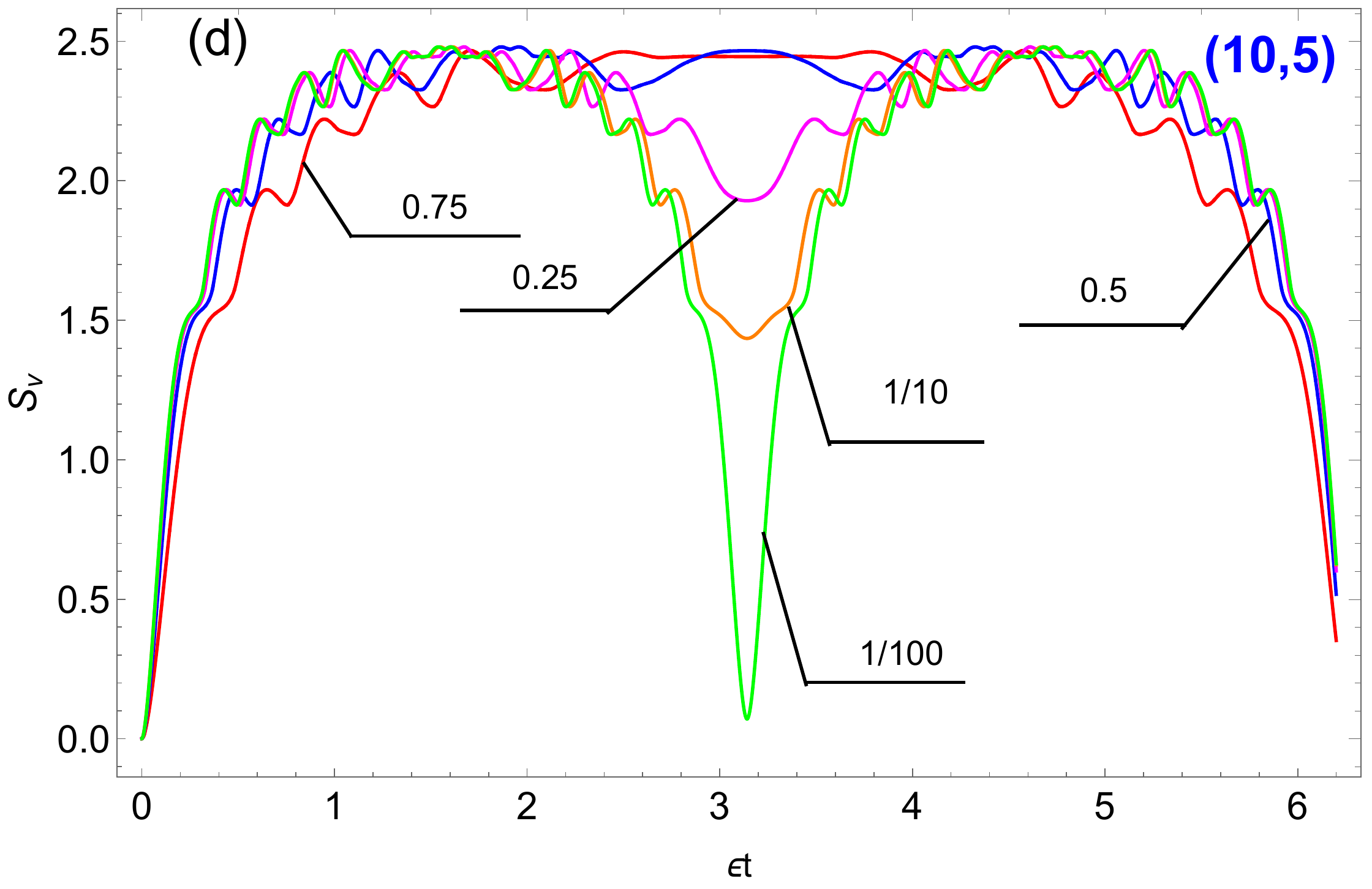}
	\captionof{figure}{\sf (color online)  The dynamics of the quantum entanglement for different values of $ \sin\theta=(0.01,0.1,0.25,0.5,0.75)$ and   quantum states $ (n,m) $.\label{fig4}}
\end{figure}
In {\bf \color{blue}\sf{Figure}} \ref{fig4}, we plot the dynamics of entanglement for the quantum states $ (0,3), (3,3), (5,5), (10,5)$ and for several choices of $ \sin\theta$. As expected, the entanglement exhibits an $ 2\pi- $periodic behavior and becomes $ \pi- $periodic as $\theta\rightarrow 0$. We mention also, that the fast generation of the maximal entanglement requires the resonance $ R\rightarrow 1$ and a time $ t\sim \epsilon^{-1} $. In contrast, the death of entanglement will be rapid for the resonance case, which is consistent with the results of position-velocity  coupling, i.e. $ x_{1}p_{2} $ obtained by Makarov \cite{4}. As a result, rapidly generating entanglement and maintaining it during the period requires  small but not too small values of $\theta $ (e.g. $ \sin\theta\sim 0.1 $). An interesting remark, is that the maximum of the entanglement entropy $ S_{v,\max}$ depends only on the quantum states $ n+m $, and this value is reached during the dynamics. We observe that the physical parameters accelerate or decelerate the generation of entanglement. 

\section{Conclusion}

 We have dealt with a specific quadratic Hamiltonian based on  two magnetically coupled harmonic oscillators. First, we have diagonalized the Hamiltonian by using three canonical transformations and obtained the exact stationary wave function as well as the associated  energies. Secondly, we have derived the analytical expression of the Schmidt modes by using Schmidt decomposition. The non-Gaussian entanglement has been studied by using two quantifiers  such that von Neumann entropy and Schmidt parameter $ K $.
 
We have analyzed the obtained results by investigating the effect of state asymmetry, anisotropy and the dynamics. It is found that the generation of entanglement is possible, by choosing the physical parameters and the quantum states $ (n,m)$. Thereby, accelerate and decelerate the generation of the maximal entanglement for a given quantum state is possible by adjusting the physical parameters. In addition, we have shown that the asymmetry of the quantum state affects the entanglement. The magnetic coupling and the anisotropy effects are discussed, and it is found that the most excited states are more sensible to the magnetic coupling and the sensitivity increases as we go away from resonance.
   The  obtained results show the possibility to obtain a very large quantum entanglement probability, which can be used for example to study the magnetically coupled wave guides beam splitters.

  \begin{appendices}
\numberwithin{equation}{section}
\section{\label{appe}Appendix:  Transformation matrix $S$ }

To express the new coordinates $ \hat{Q}_{j}, \hat{P}_{j} $ in terms of the old ones
 $ \hat{x}_{j}, \hat{p}_{j} $, we  use  three canonical transformations in their matrix forms
 \begin{eqnarray}
S_{1}=\begin{pmatrix}
	\tfrac{1}{\sqrt{2}}&0  & 0 & \tfrac{1}{\sqrt{2}\omega_{2}} \\ 
	\tfrac{1}{\sqrt{2}}&0  & 0 & -\tfrac{1}{\sqrt{2}\omega_{2}} \\  
	0&\tfrac{\omega_{2}}{\sqrt{2}} & \tfrac{1}{\sqrt{2}} & 0 \\ 
	0&-\tfrac{\omega_{2}}{\sqrt{2}} & \tfrac{1}{\sqrt{2}} & 0
\end{pmatrix}, \qquad
S_{2}=\begin{pmatrix}
\sqrt[4]{ \tfrac{m_{-}}{m_{+}}} &0  & 0 & 0 \\ 
	0&\sqrt[4]{ \tfrac{m_{+}}{m_{-}}} & 0 & 0 \\  
	0&0 & \sqrt[4]{ \tfrac{m_{+}}{m_{-}}} & 0 \\ 
	0&0 & 0& \sqrt[4]{ \tfrac{m_{-}}{m_{+}}}
\end{pmatrix}
\end{eqnarray}
\begin{eqnarray}
S_{3}=\begin{pmatrix}
\cos\tfrac{\theta}{2} & \sin\tfrac{\theta}{2} & 0 & 0 \\ 
-\sin\tfrac{\theta}{2}&\cos\tfrac{\theta}{2} & 0 & 0 \\  
0&0 & \cos\tfrac{\theta}{2} & \sin\tfrac{\theta}{2} \\ 
0&0 & -\sin\tfrac{\theta}{2}& \cos\tfrac{\theta}{2}
\end{pmatrix}
\end{eqnarray}
to obtain the mapping
\begin{eqnarray}
\begin{pmatrix}
\hat{Q}_1\\ 
\hat{Q}_2\\ 
\hat{P}_1\\
\hat{P}_2 
\end{pmatrix}
=S\begin{pmatrix}
\hat{x}_1\\ 
\hat{x}_2\\ 
\hat{p}_1\\
\hat{p}_2 
\end{pmatrix}
\end{eqnarray}
where we have set $ S=S_3 S_2 S_1 $. More explicitly, one  can write 
\begin{eqnarray}
\hat{Q}_{1}&=&\tfrac{\sqrt{2}}{2} \left(\sqrt[4]{\tfrac{m_{+}}{m_{-}}} \cos\tfrac{\theta}{2}+ \sqrt[4]{\tfrac{m_{-}}{m_{+}}}\sin\tfrac{\theta}{2}\right)\hat{x}_{1}+\tfrac{\sqrt{2}}{2\omega_{2}}\left( \sqrt[4]{\tfrac{m_{+}}{m_{-}}}  \cos\tfrac{\theta}{2}- \sqrt[4]{\tfrac{m_{-}}{m_{+}}}\sin\tfrac{\theta}{2}\right)\hat{p}_{2}\\
\hat{Q}_{2}&=&\tfrac{\sqrt{2}}{2} \left(-\sqrt[4]{\tfrac{m_{+}}{m_{-}}} \sin\tfrac{\theta}{2}+ \sqrt[4]{\tfrac{m_{-}}{m_{+}}}\cos\tfrac{\theta}{2}\right)\hat{x}_{1}-\tfrac{\sqrt{2}}{2\omega_{2}}\left( \sqrt[4]{\tfrac{m_{+}}{m_{-}}}  \sin\tfrac{\theta}{2}+ \sqrt[4]{\tfrac{m_{-}}{m_{+}}}\cos\tfrac{\theta}{2}\right)\hat{p}_{2}\\
\hat{P}_{1}&=&\tfrac{\omega_{2}}{\sqrt{2}} \left(-\sqrt[4]{\tfrac{m_{+}}{m_{-}}} \sin\tfrac{\theta}{2}+ \sqrt[4]{\tfrac{m_{-}}{m_{+}}}\cos\tfrac{\theta}{2}\right)\hat{x}_{2}+\tfrac{\sqrt{2}}{2}\left( \sqrt[4]{\tfrac{m_{+}}{m_{-}}}  \sin\tfrac{\theta}{2}+ \sqrt[4]{\tfrac{m_{-}}{m_{+}}}\cos\tfrac{\theta}{2}\right)\hat{p}_{1}\\
\hat{P}_{2}&=&-\tfrac{\omega_{2}}{\sqrt{2}} \left(\sqrt[4]{\tfrac{m_{+}}{m_{-}}} \cos\tfrac{\theta}{2}+ \sqrt[4]{\tfrac{m_{-}}{m_{+}}}\sin\tfrac{\theta}{2}\right)\hat{x}_{2}+\tfrac{\sqrt{2}}{2}\left( \sqrt[4]{\tfrac{m_{+}}{m_{-}}}  \cos\tfrac{\theta}{2}- \sqrt[4]{\tfrac{m_{-}}{m_{+}}}\sin\tfrac{\theta}{2}\right)\hat{p}_{1}.
\end{eqnarray}
\end{appendices} 

\end{document}